\documentclass[
    11pt,
    letterpaper,
    reprint,
    notitlepage,
    superscriptaddress,
    aps,pra
]{revtex4-2}

\usepackage{amsmath,amssymb,amsfonts}
\usepackage{empheq}
\usepackage{physics}
\usepackage{subdepth}
\usepackage{bm}
\renewcommand{\mathbf}{\bm}
\usepackage{dsfont}
\renewcommand{\mathbb}{\mathds}
\newcommand{\ii}{\mathrm{i}}
\usepackage[svgnames,dvipsnames]{xcolor}
\usepackage{graphicx}
\definecolor{NewBlue}{rgb}{0.1, 0.1, 0.7}
\definecolor{NewRed}{rgb}{0.7, 0.1, 0.1}
\usepackage[colorlinks,
    linkcolor=Maroon,
    citecolor=NewBlue,
    urlcolor=NewRed]{hyperref}
\usepackage{cleveref}

\renewcommand{\t}[1]{\mathrm{#1}}

\newcommand{\LigoMIT}{LIGO Laboratory, Massachusetts Institute of Technology, Cambridge, MA 02139}
\newcommand{\MechMIT}{Department of Mechanical Engineering, Massachusetts Institute of Technology, Cambridge, MA 02139}

\begin{document}

\title{A Generalized Schawlow-Townes Limit}

\author{Hudson Loughlin}
\email{hudsonl@mit.edu}
\affiliation{\LigoMIT}
\author{Vivishek Sudhir}
\affiliation{\LigoMIT}
\affiliation{\MechMIT}

\date{\today}

\begin{abstract}
	We study a class of a feedback oscillators realized by a phase-insensitive amplifier in positive feedback, 
	where either the amplifier or the feedback element may determine the oscillator's linewidth.
	The spectral purity of the output of such a device originates from basic demands of quantum mechanics and causality.
	The resulting expression generalizes the Schawlow-Townes limit, which is itself one component of 
	a standard quantum limit for feedback oscillators. 
	Recently realized bad-cavity oscillators such as super-radiant lasers and solid-state masers can 
	saturate this generalized Schawlow-Townes limit.
	This limit can be surpassed through appropriate quantum engineering: for example by atomic spin squeezing in a 
	super-radiant laser.
\end{abstract}

\maketitle

\begin{figure}[t!]
	\centering
	\includegraphics[width=1\columnwidth]{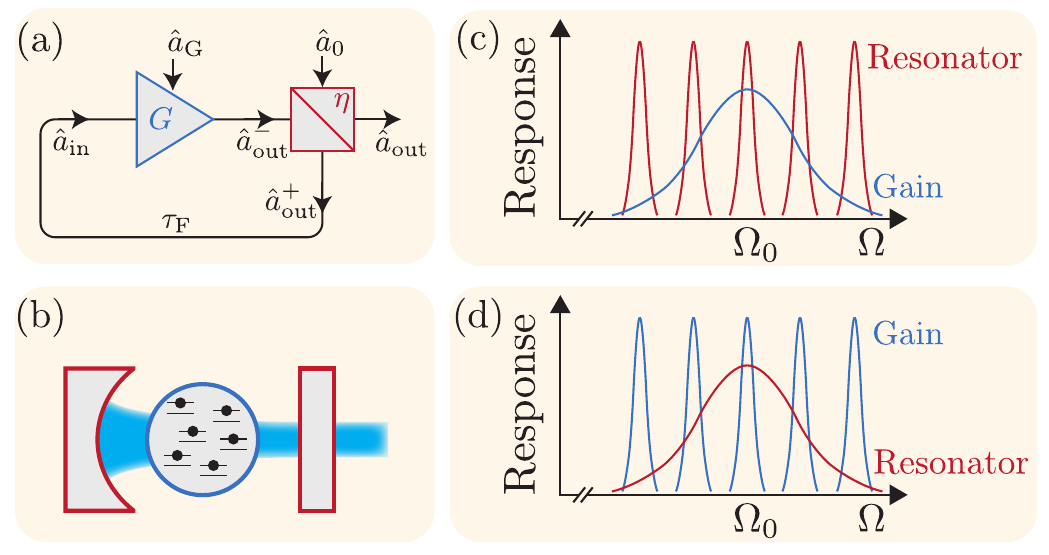}
	\caption{\label{fig:fbamp} 
	(a) Good and bad cavity feedback oscillators can be abstractly modeled as a quantum electronic or optical circuit 
	consisting of an amplifier, beam splitter, and time delay.
	(b) an example of a bad cavity feedback oscillator is a hydrogen maser or a super-radiant laser where the atomic linewidth is narrower than the cavity's. 
	(c) the gain medium and resonator response as a function of frequency for oscillators in the ``good cavity'' regime where the resonator linewidth is narrow compared to the gain medium's. 
	(d) the gain medium and resonator profiles in the ``bad cavity'' regime where the resonator linewidth is large compared with the gain medium's. 
	}
\end{figure}

\emph{Introduction.}
The spectral purity of a laser, given by the (modified) Schawlow-Townes formula \cite{SchaTow58,ScuLam67},
can be understood in a simple and general picture of an oscillator realized by an amplifier 
in positive feedback \cite{LoughSud23}.
In this picture, depicted in \cref{fig:fbamp}(a,c), the mean frequency of the oscillating output signal is dictated by the
frequency selectivity of the feedback element and its mean amplitude by the nonlinearity of the amplifier.
Quantum fluctuations in either are determined by the unavoidable quantum noises from  the amplifier and the out-coupler --- necessary elements in any feedback oscillator.
Linearized quantum noise analysis then produces the modified Schawlow-Townes formula for the linewidth \cite{LoughSud23},
\begin{equation}\label{eq:ST}
	\Gamma_\t{ST} = \frac{\hbar \Omega_0 \kappa_\text{F}^2}{2P}
\end{equation}
of the emitted signal at frequency $\Omega_0$, with power $P$, and $\kappa_\text{F}$ is the linewidth of the frequency selective
feedback element (such as a cavity in a laser). 
The Schawlow-Townes limit 
can be understood as one facet in an equal trade-off between quantum fluctuations in the
phase and amplitude of the emitted signal, i.e. a standard quantum limit for feedback oscillators. 
Knowing the precise origin of this basic result from such a minimal system-agnostic model opens the door to
the possibility of oscillators that operate beyond the Schawlow-Townes limit and to identifying the resources necessary to 
realize them. 

Lasers, and more generally feedback oscillators, need not necessarily operate in the regime where the 
feedback element is more frequency selective than the amplifier (i.e. the so-called ``good cavity'' regime
depicted in \cref{fig:fbamp}(c)).
Bad-cavity oscillators \cite{Hmaser60,KupWoe94,BenGio08,BohThom12,Oxb12,NorThomp16b,NorThomp16,Breeze18,Day24} 
(see \cref{fig:fbamp}(d)) can operate beyond 
the Schawlow-Townes limit \cite{SculBenk88}, by employing an amplifier whose linewidth is smaller 
than that of the cavity enclosing it.
Moreover, the output of such oscillators can be immune to a variety of technical noises originating in the
cavity \cite{Chen09,Meiser09,NorThomp18,Day24}, since the amplifying gain medium is well isolated from its 
environment.

We derive fundamental quantum limits on the noise performance of all feedback oscillators --- i.e. those operating
in the good as well as bad cavity regimes --- using a generalized, minimal, and system-agnostic model of such devices. 
The resulting ``generalized Schawlow-Townes limit'' applies to lasers operating
in both the good and bad cavity regimes, and can be identified as one facet of a standard quantum limit. 
This quantum limit can be surpassed, for example, by spin squeezing the atomic gain medium in a super-radiant
realization of a bad-cavity laser.

\emph{Quantum-limited feedback oscillator.}
Consider the positive feedback amplifier configuration shown in \cref{fig:fbamp}(a).
The output of a phase-insensitive amplifier with (frequency-dependent) gain $G[\Omega]$ is coupled back 
into its input after attenuation
by a factor $\eta$ and a feedback delay of $\tau_\text{F}$; the remaining fraction of the signal is coupled out of the loop to
derive the out-of-loop field $a_\t{out}$.
The equations of motion for the system are obtained by going around the loop in
\cref{fig:fbamp}(a) employing the linearity of each of the elements and of the feedback path.
The Heisenberg-picture operators in the frequency domain are then related by \cite{LoughSud23}
\begin{equation}\label{amp:eqsFT}
\begin{split}
	a_\t{out}^- [\Omega] &= G[\Omega] a_\text{in}[\Omega] + a_G[\Omega] \\
	a_\t{out}^+ [\Omega] &= -\ii\sqrt{\eta}\, a_\text{out}^- [\Omega] + \sqrt{1-\eta}\, a_0 [\Omega] \\
	a_\t{out}[\Omega] &= \sqrt{1-\eta}\, a_\text{out}^- [\Omega] +\ii\sqrt{\eta}\, a_0[\Omega] \\
	a_\t{in} [\Omega] &= e^{\ii\Omega \tau_\text{F}} a_\t{out}^+ [\Omega].
\end{split}
\end{equation}
Here, $\hat{a}_G$ ($\hat{a}_0$) represents the unavoidable noise added at the amplifier (out-coupler).
Note that the last equation is the Fourier transform of, $a_\t{in}^-(t)=a_\t{out}^+(t-\tau_\t{F})$,
which expresses the delay in the feedback path.

For oscillators operating in the good cavity regime, the spectral response of the gain medium can be approximated to be 
frequency-independent (i.e. $G[\Omega]\approx G$), since frequency selectivity of the oscillating output is enforced
by the much narrower response of the feedback loop (\cref{fig:fbamp}(c)).
For bad cavity oscillators, such a simplification is unviable.
Nevertheless, similar to a good cavity oscillator \cite{LoughSud23}, the gain will saturate to a point where ``gain = loss''. 
Since the loss is given by the out-coupler's reflectivity, $\eta$, the gain at the resonance frequency 
$\Omega_0$ will be $|G[\Omega_0]| = 1/\sqrt{\eta}$. 

Causality of the amplifier fixes the essential behavior of $G[\Omega]$.
Causality determines the \emph{minimum} phase $\phi[\Omega]=\arg G[\Omega]$ that is stipulated 
by the gain $\abs{G[\Omega]}$ \cite{Bode45,Toll56,prevedelli25}:
\begin{equation}\label{eq:kkMain}
	\phi[\omega] = 
	- \frac{1}{\pi} \text{P} \int_{-\infty}^\infty \frac{\log |G[\omega^\prime]|}{\omega^\prime - \omega} d \omega^\prime.
\end{equation}
If the amplifier has its peak gain at $\Omega_0$, then the first derivative of its magnitude response is 
zero at this frequency. Then, in the vicinity of this frequency, the magnitude response has the form
\begin{equation}\label{eq:aForm}
	|G[\Omega_0 + \omega]| = 1/\sqrt{\eta} - a \omega^2 + \mathcal{O}(\omega^3),
\end{equation}
where $\omega \equiv \Omega - \Omega_0$ and $a>0$.
Since its magnitude response is symmetric near $\Omega_0$ \cref{eq:kkMain} requires $\phi[\Omega_0] \approx 0$ such that the amplifier adds no phase on resonance. We write this relation with an approximation sign as the amplifier's magnitude response is not guaranteed to be symmetric far from resonance, and this may cause the integral in \cref{eq:kkMain} to differ slightly from zero. Then, to leading order, the (minimum) phase response is 
\begin{equation}\label{eq:phiForm}
	\phi [\Omega_0 + \omega] = \tau_\text{G} \omega + \mathcal{O}(\omega^2).
\end{equation}
Here, $\tau_\t{G}$ is some characteristic constant with the dimension of time. 
This constant has to be positive: indeed, \cref{eq:kkMain} implies that
\begin{equation}
	\frac{d\phi[\Omega_0 + \omega]}{d\omega} = \frac{1}{\pi} \text{P} \int \frac{\log \left| \frac{G[\Omega_0 + \omega]}{G[\Omega_0 + \omega^\prime]} \right|}{(\omega^\prime - \omega)^2} 
	d\omega^\prime;
\end{equation}
for an amplifier $\log \abs{G[\Omega_0]/G[\Omega_0 + \omega^\prime]} \geq 0$ (since $\abs{G[\Omega]}$ has a local maximum at $\Omega = \Omega_0$), and so $d\phi[\Omega_0 + \omega]/d\omega > 0$. 
The last inequality is strict since an amplifier must have $\abs{G[\Omega_0]/G[\Omega_0 + \omega^\prime]} > 1$ for some $\omega^\prime$. As a result the coefficient of the first-order term in the Taylor expansion of $\phi[\Omega_0 + \omega]$ will always be positive for an amplifier, 
i.e. $\tau_\text{G} > 0$ in \cref{eq:phiForm}. In the following sections, 
it will turn out that $\tau_\text{G}$ is the bare lifetime of the gain medium 
i.e. its lifetime if it were not placed inside a feedback loop. 

The input-output relation of the oscillator, obtained by solving \cref{amp:eqsFT}, can be put in the form 
\begin{equation}
	\hat{a}_\text{out} = H_0[\Omega] \hat{a}_0 + H_\text{G}[\Omega] \hat{a}_\text{G},
\end{equation}
where $H_0, H_G$ are the transfer functions from the two essential input noises 
$\hat{a}_0$ and $\hat{a}_G$ respectively, the former from the open port of the out-coupler and the latter from the
(phase-insensitive) amplifier's internal noise mode \cite{Hau62,Cav82}. The general forms of $H_{0,G}$ depend on the
amplifier's gain $G[\Omega]$. For the relevant case where the magnitude and phase of $G[\Omega]$ are given by 
\cref{eq:phiForm,eq:aForm}, and for frequency shift $\omega$ small around the resonance 
(i.e. $\abs{\omega}(\tau_\text{G} + \tau_\text{F}) \ll 1$) \footnote{
	We neglect frequency-pulling effects that arise when the amplifier and feedback element's resonances do not coincide.
}, we have
\begin{equation}
	H_0[\Omega] \approx H_\text{G}[\Omega] \approx 
	\frac{\sqrt{\eta} - 1/\sqrt{\eta}}{\ii \omega (\tau_\text{G} + \tau_\text{F})}.
\end{equation}

The linewidth of the emitted output is given in terms of the noise spectrum of its 
phase quadrature by \cite{LoughSud23},
$\Gamma = (\omega^2/2|\alpha|^2) \bar{S}_{pp}^\text{out}[\Omega_0 + \omega]$,
where $|\alpha|^2$ is the mean photon flux out of the oscillator and $\bar{S}_{pp}^\text{out}$ is 
noise spectrum of the output's phase quadrature, which is related to the spectrum of the input noises as 
\begin{equation}
\begin{split}
	\bar{S}_{pp}^\text{out}[\Omega] =& |H_0[\Omega]|^2 \bar{S}_{pp}^0[\Omega] + |H_\text{G}[\Omega]|^2 \bar{S}_{pp}^\text{G}[\Omega] \\ &\qquad- 2 \text{Re}\left[ H_0[\Omega]^* H_\text{G}[\Omega] \bar{S}_{pp}^{0,\text{G}}[\Omega] \right].
\end{split}
\end{equation}
Assuming the noise modes are in thermal states, i.e. 
$\bar{S}_{pp}^0[\Omega] = \tfrac{1}{2}+\bar{n}_0$, $\bar{S}_{pp}^\text{G}[\Omega] = \tfrac{1}{2} + \bar{n}_\t{G}$, and $\bar{S}_{pp}^{0,\text{G}}[\Omega] = 0$, 
we have
\begin{equation}\label{eq:fbOscStSpectra}
	\bar{S}_{pp}^\text{out}[\Omega] = \frac{(\sqrt{\eta} - 1/\sqrt{\eta})^2}{\omega^2 
	(\tau_\text{G} + \tau_\text{F})^2}\left( 1 + 2 \bar{n}_\text{th} \right),
\end{equation}
where $\bar{n}_\t{th} = (\bar{n}_0 + \bar{n}_G)/2$ is the mean thermal occupation of the two noise modes.
Using this expression for the phase noise spectrum, the linewidth takes the simple form
\begin{equation}\label{eq:badCavitySt}
	\Gamma = \frac{\hbar \Omega_0}{2P} 
	\frac{1 + 2\bar{n}_\text{th}}{(\kappa_\t{F}^{-1} + \kappa_\t{G}^{-1})^2}.
\end{equation}
Here, $P = \hbar\Omega_0 \abs{\alpha}^2$ is the mean power of the output; the linewidths of the feedback loop 
and that of the gain medium are $\kappa_\text{F,G} \equiv (1-\eta)\tau_\t{F,G}^{-1}$ 
(we assumed that $1-\eta \ll 1$ and the feedback loop has high quality factor). 

The above expression for the linewidth applies equally to good or bad cavity oscillators depending on whether
$\kappa_\text{F}$ is greater than or less than $\kappa_\text{G}$. Indeed, the linewidths of the amplifier and feedback loop 
``add in parallel'', so that the resulting oscillator's output linewidth is set by the narrower of the gain medium and 
the feedback loop.
For example, when the gain is broadband, i.e. $\kappa_\text{G} \gg \kappa_\text{F}$, the above expression reduces to the usual
Schawlow-Townes formula, $\Gamma_\text{ST} \approx (\hbar \Omega_0 \kappa_\text{F}^2/P) (\tfrac{1}{2} + \bar{n}_\text{th})$,
that applies to good-cavity lasers. For this reason, the expression in \cref{eq:badCavitySt} will henceforth be referred
to as the generalized Schawlow-Townes linewidth, and denoted $\Gamma_\t{GST}$.
Our result also agrees with calculations specific to more general 
lasers \cite{Haken64,Haken66}. 
Importantly, our expression applies even more broadly since our model is agnostic to the specific amplifying mechanism 
and only relies on the barest essential requirements imposed by the principles of causality and quantum mechanics.

\emph{Standard quantum limit for feedback oscillators.} The generalized Schawlow-Townes linewidth for feedback 
oscillators is one facet of a standard quantum limit (SQL). 
To wit, when no quantum enhancement is employed, the uncertainty principle for the output field, 
quantified in terms of the product of the noise spectra
of the quadratures of the output can be bounded as \cite[\S IV.C]{Ding24}
$\bar{S}_{qq}^\text{out}[\Omega]\bar{S}_{pp}^\text{out}[\Omega] \geq 
\tfrac{1}{4}(|H_0[\Omega]|^2 + |H_\text{G}[\Omega]|^2)^2$. Explicitly, 
\begin{equation}\label{eq:fbOscSql}
	\bar{S}_{qq}^\text{out}[\Omega]\bar{S}_{pp}^\text{out}[\Omega] 
	\gtrsim \left( \omega (\kappa_\t{F}^{-1} + \kappa_\t{G}^{-1}) \right)^{-4};
\end{equation}
this inequality can be interpreted as an SQL for generalized feedback oscillators.
(Here the approximation is valid for $\eta - 1 \ll 1$.)
Comparing \cref{eq:fbOscSql} to \cref{eq:fbOscStSpectra}, we see that the Schawlow-Townes limit arises from the 
SQL in the case that the uncertainty in either quadrature equally splits the lower bound. 

The SQL can be evaded by engineering an asymmetry between the two quadratures, or by modifying the oscillator's dynamics and hence its transfer functions. For instance, if the in-coupled and amplifier's noise modes are phase-squeezed 
with squeezing parameters $r_0$ and $r_\text{G}$ respectively, 
the oscillator has a linewidth given by $\Gamma = \Gamma_\text{GST} ( e^{-2r_0} + e^{-2r_\text{G}})/2$.

\textit{Application to super-radiant lasers.} In what follows, we show that super-radiant lasers 
can saturate the ``bad-cavity'' Schawlow-Townes linewidth limit and explore how the linewidth of such a laser can be narrowed beyond this limit by squeezing or entangling its atoms and/or incident electromagnetic field.

A super-radiant laser can be modeled as a collection of $N$ spin-1/2 systems, almost entirely in their 
excited state, interacting with an optical cavity mode such that the atomic transition is much narrower than that of
the cavity. The effective spin-$N/2$ system, described by the collective angular momentum 
operators $\{\hat{S}_x$, $\hat{S}_y$, $\hat{S}_z\}$, resonantly interacts with a single cavity mode via the 
Jaynes-Cummings Hamiltonian \cite{Meiser09,Bohnet18}, $\hat{H}_\t{int} = \ii \hbar g ( \hat{a} \hat{S}_+ 
- \hat{a}^\dagger \hat{S}_- )$.
Here, 
$\hat{S}_\pm = \hat{S}_x \pm \ii \hat{S}_y$ are the spin raising and lower operators, and $\hat{a}^\dagger,\hat{a}$ are 
those of the cavity mode. 
In a super-radiant laser, the atoms are almost entirely in the excited state. It is then useful to use the Holstein-Primakoff mapping to relate the atomic spin operators to an effective bosonic mode $\hat{b}$ by \cite{Holstein40,Persico75,Klein91},
$\hat{S}_+ = \sqrt{2N}\sqrt{1-\hat{b}^\dagger\hat{b}/(2N)} \, \hat{b} = \hat{S}_-^\dagger$.
Since the atoms are almost all in the excited state, by assumption, $\hat{b}^\dagger\hat{b}/(2N) \ll 1$, in which 
case the interaction Hamiltonian $\hat{H}_\t{int}$ describes 
a non-degenerate parametric amplifier. 

To describe fluctuations in the atomic and cavity sub-systems, we linearize the interaction by taking $\hat{a} = \alpha + \delta \hat{a}$ and $\hat{b} = \beta + \delta \hat{b}$ where $\delta \hat{a}$ and $\delta \hat{b}$ represent small quantum fluctuations on top of the large mean field values $\alpha$ and $\beta$. Working to leading order in $\delta\hat{a}$  and $\delta\hat{b}$, we have $\hat{S}_+ \approx \sqrt{2N-|\beta|^2}\,\hat{b}$ etc. 
A consistent accounting of the fluctuations requires a description of the spontaneous emission of the atoms and 
fluctuations in the field entering through the laser's partially reflective mirror. 
We do so using input-output theory \cite{Collett1984,Gardiner85} and find that the super-radiant laser evolves according to
\begin{equation}\label{eq:superRadInOut}
\begin{split}
	\dot{\hat{a}} &= -\frac{\kappa_\text{F}}{2} \hat{a} - \sqrt{2N - |\beta|^2}\,g \hat{b}^\dagger - \sqrt{\kappa_\text{F}} \hat{a}_\text{in} \\
	\dot{\hat{b}}^\dagger &= -\frac{\kappa_\text{G}}{2} \hat{b}^\dagger - \sqrt{2N- |\beta|^2}\,g \hat{a} - \sqrt{\kappa_\text{G}} \hat{b}_\text{in}^\dagger \\
	\hat{a}_\text{out} &= \sqrt{\kappa_\text{F}} \, \hat{a} + \hat{a}_\text{in}.
\end{split}
\end{equation}

The classical steady-state of the laser follows from demanding that the expectation values $\alpha,\beta$ do not
evolve in time. This gives
\begin{equation}\label{eq:superRadMeanField}
\begin{split}
	0 &= \beta \left( 2 N - \frac{\kappa_\text{F} \kappa_\text{G}}{4 g^2} - |\beta|^2 \right) \\
	\alpha &= -\frac{2 g}{\kappa_\text{F}}\sqrt{2N - |\beta|^2} \, \beta^*,
\end{split}
\end{equation}
so the system has a lasing threshold at $C \equiv 8g^2N/(\kappa_F \kappa_G) = 1$; here $C$ is the
collective cooperativity.
When $C<1$, the steady-state amplitude is given by $\beta = 0$ and the system has no macroscopic output. 
When $C>1$, the steady-state amplitudes are given by
\begin{equation}\label{eq:steadyStateField}
\begin{split}
	|\beta|^2 = 2 N(1-C^{-1}), \quad
	|\alpha| = \sqrt{\kappa_\text{G}/\kappa_\text{F}} \, |\beta|,
\end{split}
\end{equation}
implying that the laser emits a mean photon flux $\abs{\alpha_\t{out}}^2 = 2N \kappa_G (1 - C^{-1})$.

Fluctuations on top of the steady-state flux can be computed from \cref{eq:superRadInOut} by linearizing about
the steady-state in \cref{eq:steadyStateField}.
Fourier transforming and solving for the output field mode fluctuations $\delta \hat{a}_\text{out}$ gives
\begin{equation}
\begin{split}
	\delta\hat{a}_\text{out}[\omega] \approx 
	-\ii \frac{\delta\hat{a}_\text{in}[\omega] + \delta\hat{b}_\text{in}^\dagger[\omega]}
	{(\kappa_\text{F}^{-1} + \kappa_\text{G}^{-1})\omega} 
\end{split}
\end{equation}
This expression is approximated for frequency detunings small about resonance, which is the regime relevant for determining
the linewidth of the output.
The spectrum of the oscillator's output amplitude and phase quadrature fluctuations then follow:
\begin{equation}\label{eq:SppOutSuperRad}
	\bar{S}_{xx}^\text{out} = 
	\frac{\bar{S}_{xx}^a \pm 2 \bar{S}_{xx}^{ab} + \bar{S}_{xx}^b }{[(\kappa_\t{F}^{-1}+\kappa_\t{G}^{-1})\omega]^2},
\end{equation}
where all spectra should be evaluated at the frequency $\Omega_0 + \omega$ and $x \in {q,p}$ is the amplitude or phase quadrature, related to the lowering operator by $\hat{q} \equiv (\hat{a}^\dagger + \hat{a})/\sqrt{2}$ 
and $\hat{p} \equiv i(\hat{a}^\dagger - \hat{a})/\sqrt{2}$; here, the $\pm$ signs correspond to $q,p$.
In particular, the spectrum of phase fluctuations gives the linewidth; 
assuming the electromagnetic and atomic incident modes are in thermal states, the oscillator's linewidth is
\begin{equation}\label{eq:generalizedStSuperRad}
\begin{split}
	\Gamma &= \frac{\hbar \Omega_0}{2P} 
	\frac{1 + 2 n_\t{th}}{(\kappa_\text{F}^{-1} + \kappa_\text{G}^{-1})^2} 
\end{split}
\end{equation}
where $\bar{n}_\t{th}$ is the average thermal population of the atomic and field sub-systems, and 
$P=\hbar\Omega\abs{\alpha_\t{out}}^2$ is the mean output power.
Comparing \cref{eq:generalizedStSuperRad} with \cref{eq:badCavitySt}, we see that the super-radiant laser saturates the ``bad-cavity'' Schawlow-Townes linewidth limit.

\textit{Quantum-enhanced super-radiant laser.} A super-radiant laser without quantum-engineered inputs and outputs has a linewidth limited by \cref{eq:generalizedStSuperRad}. As apparent from \cref{eq:SppOutSuperRad}, the linewidth of a super-radiant laser can be suppressed beyond the Schawlow-Townes limit by either squeezing the atomic and/or cavity 
field, or correlating the in-coupled noise modes $\hat{a}_\text{in}$ and $\hat{b}_\text{in}$.
While the optical mode $\hat{a}_\text{in}$ represents a single spatial mode, defined by the cavity, and can be readily squeezed, the mode $\hat{b}_\text{in}$ represents the collection of all spatial modes the atoms emit photons 
into besides the cavity mode. This huge collection of modes cannot be readily squeezed with existing technology.

Instead, we consider directly squeezing the collective atomic 
excitation \cite{Meyer01,Leibfried04,AppPolz09,LerVul10,GrosOber10,SewMitch12,HosKas16,Miller24}. 
This can be modeled by adding a spin squeezing term to the interaction Hamiltonian, which
can either take the form of a one-axis twisting interaction $\hbar \chi_1(\hat{S}_x - \hat{S}_y)^2$
\cite{Wineland92,KitUeda93,Meyer01,Leibfried04}, 
or a two-axis twisting interaction $(\ii \hbar \chi_2/2)(\hat{S}_-^2 - \hat{S}_+^2)$ \cite{KitUeda93,Liu11,Miller24,Luo24}. 
In either case, after making the Holstein-Primakoff approximation the interaction Hamiltonian in the rotating frame is
\begin{equation}\label{eq:spinSqzHamiltonian}
	\hat{H}_\t{int} = \ii \hbar \sqrt{2N} \, g \left(\hat{a}\hat{b} - \hat{a}^\dagger \hat{b}^\dagger\right) + \ii \hbar N \chi \left(\hat{b}^2 - \hat{b}^{\dagger 2} \right).
\end{equation}
Here $\chi$ is either $\chi_{1,2}$ depending on whether the interaction is one or two-axis twisting.

Proceeding as in the case of a conventional super-radiant laser, we find that the system has a lasing 
threshold when $C = (1+s)^{-1}$, where $s \equiv \kappa_\text{F} \chi / (2g^2)$ is a dimensionless factor characterizing the strength of the spin-squeezing interaction \footnote{
	Note that $2s = [N\chi]/[(\sqrt{2N}g)^2/(\kappa_F/2)]$ is the ratio of the rate of spin squeezing generation to
	the rate at which atomic excitation leaves the cavity as optical photons, thus it plays the role of the squeezing
	factor in this problem.
}.
As before, we linearize about the mean-field values in order to find the spectra of the oscillators' quantum fluctuations. 
The crucial upshot of the result is that, in contrast to the case with no spin squeezing, we now have an asymmetry between the amplitude and phase quadratures. 
In the simple case of Fourier frequencies close to resonance and when spin-squeezing is weak 
(see \Cref{app:spinSqzDetails} for exact expressions), 
the output amplitude and phase quadratures are given by
\begin{equation}
	\delta \hat{q}_a^\text{out}[\omega] = -\ii 
	\frac{\delta \hat{q}_a^\text{in}[\omega] -\delta  \hat{q}_b^\text{in}[\omega]}
	{(\kappa_\text{F}^{-1} + \kappa_\text{G}^{-1} ) \omega},
\end{equation}
\begin{equation}\label{eq:poutSuperRad}
	\delta \hat{p}_a^\text{out}[\omega] = 
	\frac{\delta \hat{p}_a^\text{in}[\omega]  + \delta  \hat{p}_b^\text{in}[\omega]}
	{s (C - 2) \kappa_\text{F} + \ii (\kappa_\text{F}^{-1}+\kappa_\text{G}^{-1}) \omega}.
\end{equation}
Notably, the transfer function in this expression no longer has a pole at $\omega = 0$, allowing the phase-quadrature fluctuations to be substantially reduced below the generalized Schawlow-Townes limit.
From \cref{eq:poutSuperRad}, the linewidth of the spin-squeezed super-radiant laser's output is given by 
(see \Cref{app:spinSqzLinewidth} for details)
\begin{equation}
	\Gamma \approx \Gamma_\t{GST} \left[ 1 -  
	\left( \frac{\Gamma_\text{GST}}{\gamma s^2} \right)^{-1}
	\tan^{-1} \left( \frac{\Gamma_\text{GST}}{\gamma s^2}\right) \right]^{1/2},
\end{equation}
where $\gamma = [4 \ln(2)/\pi^2] (2-C)^2 /(\kappa_\text{F}^{-1} + \kappa_\text{G}^{-1})$ characterizes the linewidth reduction due to spin-squeezing. Clearly, the linewidth narrows as the level of spin-squeezing increases.

\begin{figure}[t!]
	\centering
	\includegraphics[width=1\columnwidth]{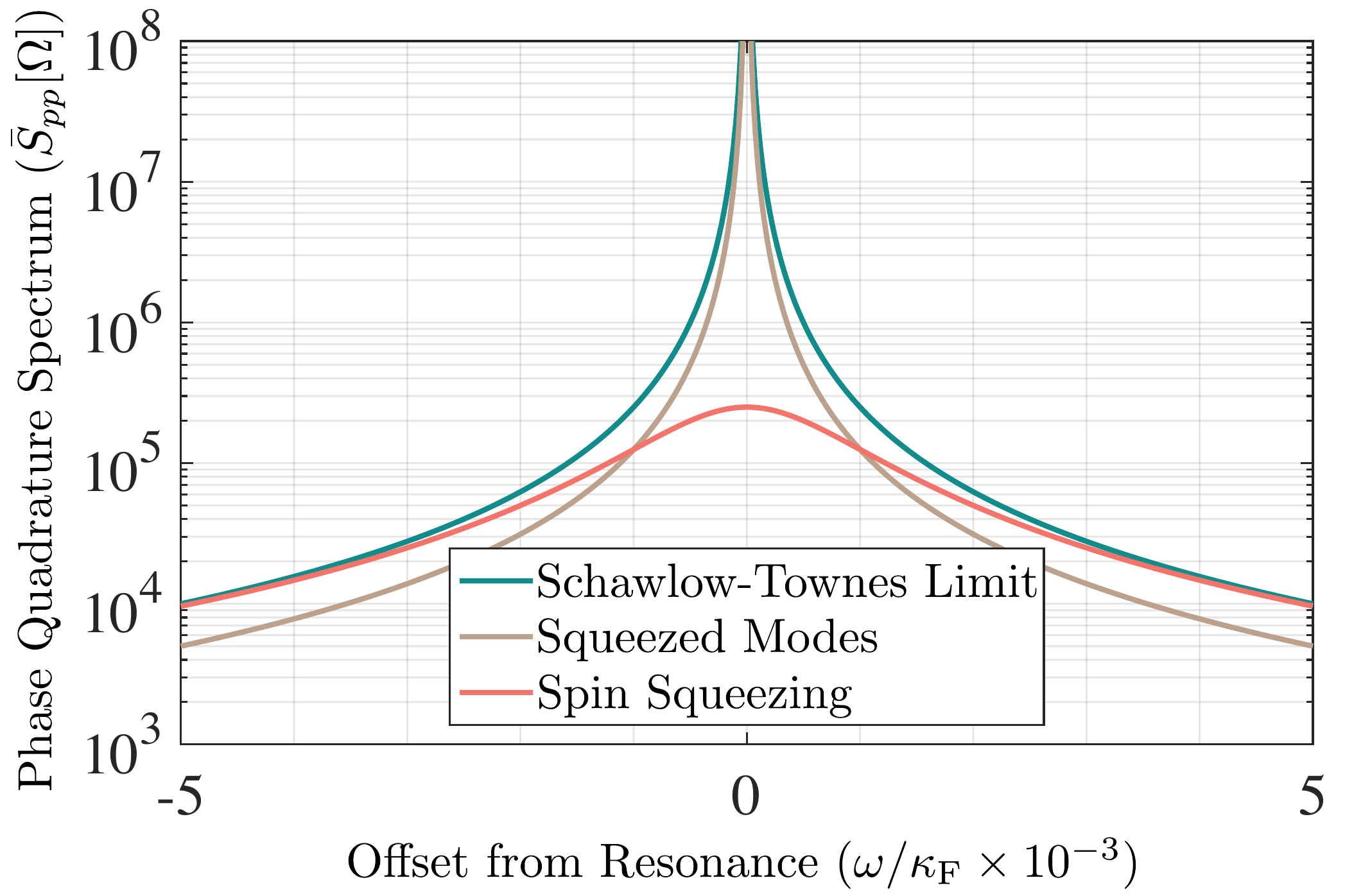}
	\caption{\label{fig:fbSpec} The phase quadrature spectra for super-radiant lasers. All curves take $\kappa_\text{F} = \kappa_\text{G}$. The ``Schawlow-Townes'' curve plots the generalized Schawlow-Townes limit for such an oscillator, 
	and the ``squeezed modes'' curve plots the bound for an oscillator with highly squeezed vacuum light injected into its input/output coupler. 
	The ``Spin squeezing'' curve plots the spectra for a spin-squeezed super-radiant laser with the spin-squeezing factor $s|C-2| = 0.002$. 
	}
\end{figure}

\Cref{fig:fbSpec} shows the spectra of several types of quantum-engineered super-radiant lasers. Without quantum 
enhancement, these lasers' phase quadrature fluctuations are limited by the generalized Schawlow-Townes limit. 
This limit can be evaded by injecting squeezed light into the device's input/output coupler, however, such improvement
is limited to at most a halving of the frequency noise.
Further improvement requires modifying the system's underlying dynamics, for example by spin-squeezing the
atomic excitation that produces gain.

\textit{Conclusion.} We derived a generalized Schawlow-Townes limit, applicable to feedback oscillators that operate
in the ``good'' or ``bad'' cavity regimes, including, but not limited to, lasers and masers.
This limit is a consequence of necessary consistency conditions --- arising from causality and quantum mechanics ---
that any feedback oscillator must obey.
Having identified the quantum origins of this bound, we find that it is not a fundamental limit, but a 
standard quantum limit (SQL). A super-radiant laser can in principle saturate this SQL, and operate at the generalized
Schawlow-Townes limit.
Appropriate quantum engineering can evade the SQL;
in the case of a super-radiant laser, it can be evaded by atomic spin squeezing. Given the promise of 
``bad'' cavity lasers, and in particular super-radiant lasers, as an avenue for realizing local oscillators that 
are not limited by technical noises, our results
establish fundamental benchmarks to gauge their performance against, and a route for their continued improvement
beyond the standard quantum limit.

\bibliographystyle{apsrev4-2}
\bibliography{ref_feedback_oscillator}

\clearpage
\appendix

\section{Details of a Super-Radiant Laser with Spin Squeezing}\label{app:spinSqzDetails}

From the Hamiltonian in \cref{eq:spinSqzHamiltonian}, we find that the equations of motion for a spin-squeezed super-radiant laser are given by
\begin{equation}\label{eq:spinSqzInOut}
\begin{split}
	\partial_t \hat{a} =& -\frac{\kappa_\text{F}}{2} \hat{a} - \sqrt{2N - |\beta|^2}\,g \hat{b}^\dagger - \sqrt{\kappa_\text{F}} \hat{a}_\text{in} \\
	\partial_t \hat{b}^\dagger =& -\frac{\kappa_\text{G}}{2} \hat{b}^\dagger - \sqrt{2N- |\beta|^2}\,g \hat{a} \\& + 2 \chi \left( N - |\beta|^2 \right) \hat{b} - \sqrt{\kappa_\text{G}} \hat{b}_\text{in} \\
	\hat{a}_\text{out} =& \sqrt{\kappa_\text{F}} \, \hat{a} + \hat{a}_\text{in}.
\end{split}
\end{equation}
Assuming $\beta \in \mathbb{R}$ such that the macroscopic field is aligned with the oscillator's 
amplitude quadrature \footnote{
In the spin-squeezed case, this assumption is necessary since the drive that determines the squeezing angle implies
a preferred quadrature. Note that classical noise in this drive needs to be considered in any practical implementation 
of this scheme.}, 
and making the Holstein-Primakoff approximation that $|\beta|^2 \ll N$, we find that the steady-state mean-field amplitudes are the solutions to the equations
\begin{equation}
\begin{split}
	\alpha &= -\frac{2g}{\kappa_\text{F}} \sqrt{2N - |\beta|^2} \beta \\
	0 &= \beta \left[ 2N \left(1 + s - C^{-1} \right) -  \left(1 + 2s \right) |\beta|^2 \right].
\end{split}
\end{equation}
We note that these equations reduce to \cref{eq:superRadMeanField} in the case that $\chi = 0$ and there is no two-axis twisting interaction. We see that as before, the system has a lasing transition below which $\beta = 0$ and the system has no mean output and above which the system has a macroscopic output given by
\begin{equation}\label{eq:meanFieldSpinSqz}
\begin{split}
	|\beta|^2 &= 2N\frac{ 1 + s - C^{-1}}{1 + 2s}\\
	|\alpha| &= \sqrt{\frac{\kappa_\text{G}(1+s C)}{\kappa_\text{F} (1+2s)}} \, |\beta|.
\end{split}
\end{equation}
As expected, these equations reduce to \cref{eq:steadyStateField} in the absence of a two-axis twisting interaction.

Using \cref{eq:meanFieldSpinSqz}, \cref{eq:spinSqzInOut} tells us that the system's quantum fluctuations are governed by
\begin{equation}
\begin{split}
	\partial_t \, \delta \hat{a} =& -\frac{\kappa_\text{F}}{2} \delta \hat{a} - \sqrt{2N} \, g \sqrt{\frac{s+C^{-1}}{1+2s}} \, \delta\hat{b}^\dagger - \sqrt{\kappa_\text{F}} \delta\hat{a}_\text{in} \\
	\partial_t \, \delta \hat{b}^\dagger =& -\frac{\kappa_\text{G}}{2} \delta \hat{b}^\dagger - \sqrt{2N} \, g \sqrt{\frac{s+C^{-1}}{1+2s}} \, \delta \hat{a} \\& + \frac{\kappa_\text{G} s (2-C)}{2(1+2s)} \delta\hat{b} - \sqrt{\kappa_\text{G}} \,\delta\hat{b}_\text{in}^\dagger \\
	\delta\hat{a}_\text{out} =& \sqrt{\kappa_\text{F}} \, \delta\hat{a} + \delta\hat{a}_\text{in}.
\end{split}
\end{equation}
We emphasize that we have used the Holstein-Primakoff approximation here to neglect fluctuation terms inside the quantity $\sqrt{2N - \hat{b}^\dagger \hat{b}}$ and replace this with $\sqrt{2N - |\beta|^2}$. As a result, these equations are only valid for $|\beta|^2 \ll 2N$.
Defining the amplitude and phase quadratures for mode $i$ as $\hat{q}_i \equiv (\hat{a}_i^\dagger + \hat{a}_i)/\sqrt{2}$ and $\hat{p}_i \equiv \ii(\hat{a}_i^\dagger - \hat{a}_i^\dagger)/\sqrt{2}$, Fourier transforming, we find that the oscillator's output amplitude quadrature is given by
\begin{equation}
\begin{split}
	&\delta \hat{q}_a^\text{out}[\omega] = \frac{-\ii \kappa_\text{F} \kappa_\text{G} \left( 1 + C s \right)}{(\kappa_\text{F}(1 + 2s) + \kappa_\text{G} (1 + C s)) \omega} \delta\hat{q}_a^\text{in}[\omega] \\& \quad + \frac{\ii \kappa_\text{F} \kappa_\text{G} (1+2s) \sqrt{(1+2s)(1+Cs)}}{(\kappa_\text{F}(1 + 2s) + \kappa_\text{G} (1 + C s)) \omega} \delta\hat{q}_b^\text{in}[\omega]
\end{split}
\end{equation}
As in the main text, we have neglected non-leading terms in $\omega$ since we are primarily interested in the oscillator's noise near resonance. Similarly, the oscillator's output phase quadrature is given by
\begin{widetext}
\begin{equation}
\begin{split}
	\delta\hat{p}_a^\text{out}[\omega] &= \frac{(1+2s)\kappa_\text{F} \kappa_\text{G} + \ii \omega(1 +(4 - C) s \kappa_\text{G})}{\ii \omega (\kappa_\text{F} + \kappa_\text{G}) + s ((C-2) \kappa_\text{F} \kappa_\text{G} + 2 \ii \kappa_\text{F} \omega - \ii (C-4) \kappa_\text{G} \omega) } \delta\hat{p}_a^\text{in}[\omega] \\
	&+ \frac{(1+2s) \sqrt{\frac{1+C s}{1 + 2 s}} \, \kappa_\text{F} \kappa_\text{G} }{ \ii \omega (\kappa_\text{F} + \kappa_\text{G}) + s ((C-2) \kappa_\text{F} \kappa_\text{G} + 2 \ii \kappa_\text{F} \omega - \ii (C-4) \kappa_\text{G} \omega) } \delta\hat{p}_b^\text{in}[\omega],
\end{split}
\end{equation}
\end{widetext}
where we have neglected quadratic terms in $\omega$.

Assuming that $\chi$ and $\omega$ are both small parameters, we can simplify the preceding equations to
\begin{equation}\label{eq:spinSqzOutApprox}
\begin{split}
	\delta\hat{q}_a^\text{out}[\omega] &= \frac{-\ii \kappa_\text{F} \kappa_\text{G}}{\left(\kappa_\text{F} + \kappa_\text{G}\right) \omega} \left(\delta\hat{q}_a^\text{in}[\omega] - \delta\hat{q}_b^\text{in}[\omega] \right)  \\
	\delta\hat{p}_a^\text{out}[\omega] &= \frac{\kappa_\text{F} \kappa_\text{G} \left( \delta\hat{p}_a^\text{in}[\omega]  + \delta\hat{p}_b^\text{in}[\omega] \right)}{s (C-2) \kappa_\text{F} \kappa_\text{G} + \ii \omega (\kappa_\text{F} + \kappa_\text{G})} .
\end{split}
\end{equation}

\section{The Linewidth of a Super-Radiant Laser with Spin Squeezing}\label{app:spinSqzLinewidth}

For an oscillator with a flat frequency noise spectrum, the linewidth is given in terms of the frequency noise spectrum by $\Gamma = \bar{S}_{\dot{\varphi} \dot{\varphi}}/(2\pi)$ \cite{Domenico10}. For an oscillator with a frequency noise spectrum that varies with the frequency offset from resonance, the relation between the frequency spectrum and linewidth is generally more complicated, but is given approximately by \cite{Domenico10}
\begin{equation}
	\Gamma = \sqrt{8 \ln(2) A},
\end{equation}
where
\begin{equation}\label{eq:Aeq}
	A := \int_0^{\Omega_\text{cut}} \frac{\omega^2}{4 \pi^2 |\alpha|^2} \bar{S}_{pp}^\text{out}[\omega] \frac{d \omega}{2\pi}
\end{equation}
and $\Omega_\text{cut}$ is determined by
\begin{equation}\label{eq:omegaCut}
	\frac{16 \ln(2) |\alpha|^2}{\pi \Omega_\text{cut}} = \bar{S}_{pp}^\text{out}[\Omega_\text{cut}].
\end{equation}

To arrive at a relatively simple expression for the linewidth, we assume that the amount of spin-squeezing is sufficiently small that we can approximate $\bar{S}_{pp}^\text{out}[\Omega_\text{cut}]$ by $\bar{S}_{pp}^\text{out}[\Omega_\text{cut}]|_{s=0}$ in \cref{eq:omegaCut}. Making this approximation and evaluating the integral in \cref{eq:Aeq} using $\bar{S}_{pp}^\text{out}[\omega]$ from \cref{eq:spinSqzOutApprox}, we find
\begin{equation}
	\Gamma \approx \sqrt{\Gamma_\text{GST}^2 -  \gamma \Gamma_\text{GST} s^2 \, \tan^{-1} \left( \frac{\Gamma_\text{GST}}{\gamma s^2} \right)},
\end{equation}
where 
\begin{equation}
	\gamma := \frac{4 \ln(2) (C-2)^2 \kappa_\text{F} \kappa_\text{G}}{\pi^2 (\kappa_\text{F} + \kappa_\text{G})}
\end{equation}
characterizes the linewidth reduction due to the spin-squeezing interaction. As expected, the linewidth reduces to the Schawlow-Townes linewidth in the absence of a spin squeezing interaction, i.e. for $s = 0$.

\end{document}